\documentclass[11pt]{article}
\usepackage{float}
\usepackage{amssymb, amsmath, amsthm}
\usepackage{mathrsfs}
\usepackage{tabulary}
\usepackage{rotating}
\usepackage{enumerate}
\usepackage{setspace}
\usepackage{multirow}
\usepackage{paralist}
\usepackage{listings}
\usepackage[T1]{fontenc}
\usepackage[utf8]{inputenc}
\usepackage[english]{babel}
\usepackage{authblk}
\usepackage{etoolbox}
\usepackage{longtable}
\usepackage[numbers,round,sort&compress]{natbib}\usepackage[usenames,dvipsnames]{color}
\usepackage{bibentry}

\newcommand{\E}{\mathsf{E}}

\newcommand{\dd}{\mathrm{d}}

\newcommand\BibTeX{{\rmfamily B\kern-.05em \textsc{i\kern-.025em b}\kern-.08em
T\kern-.1667em\lower.7ex\hbox{E}\kern-.125emX}}

\usepackage{caption}

\IfFileExists{upquote.sty}{\usepackage{upquote}}{}

\usepackage{hyperref}
\usepackage{subcaption}

\usepackage[hmargin=2.54cm,vmargin=2.54cm]{geometry}

\title{When effects cannot be estimated: redefining estimands to understand the effects of naloxone access laws} 

\vspace{1cm}
\date{}

\newcommand{\indep}{\mbox{$\perp\!\!\!\perp$}}

\author[1]{Kara E. Rudolph\thanks{Corresponding author: \\
kr2854@cumc.columbia.edu \\ 722 W. 168th St, NY, NY 10032 \\ tel. +12123422926 }}
\author[1]{Catherine Gimbrone}
\author[2]{Ellicott C. Matthay}
\author[3]{Iv\'an D\'iaz}
\author[4]{Corey S. Davis}
\author[1]{Katherine Keyes}
\author[5]{Magdalena Cerd\'a}
\affil[1]{\footnotesize Department of Epidemiology, Mailman School of Public Health, Columbia University, New York, New York}
\affil[2]{\footnotesize Center for Health and Community, School of Medicine, University of California, San Francisco}
\affil[3]{\small Division of Biostatistics, Department of Population
  Health Sciences, Weill Cornell Medicine, New York, New York}
\affil[4]{\footnotesize Network for Public Health Law, Los Angeles, California}
\affil[5]{\footnotesize Center for Opioid Epidemiology and Policy, Department of Population Health, School of Medicine, New York University, New York, New York}
\date{}

\begin{document}
\maketitle

\doublespacing
\begin{abstract}
Violations of the positivity assumption (also called the common support condition) challenge health policy research, and can result in significant bias, large variance, and invalid inference. We define positivity in the single- and multiple-timepoint (i.e., longitudinal) health policy evaluation setting, and discuss real-world threats to positivity. 
 We show empirical evidence of the practical positivity violations that can result when attempting to estimate effects of health policies (in this case, Naloxone Access Laws). In such scenarios, an alternative is to estimate the effect of a shift in law enactment (e.g., the effect if enactment had been delayed by some number of years). Such an effect corresponds to what is called a modified treatment policy, and dramatically weakens the required positivity assumption, thereby offering a means to estimate policy effects even in scenarios with serious positivity problems. We apply the approach to define and estimate longitudinal effects of Naloxone Access Laws on opioid overdose rates.\\
\end{abstract}

\newpage

\section*{Introduction}
Estimating the health effects of policies, such as drug- and firearm-related laws, has become increasingly common in public health.\cite{rudolph2015association, hamilton2021good,santaella2016we} For example, the opioid overdose crisis\cite{hedegaard2017drug,wilson2020drug,hedegaarddrug,cdccovid} has prompted the development and enactment of numerous, novel harm-reduction laws.\cite{schuler2020state} 
 It is important to understand the effectiveness of these laws 
 at achieving their goals. As a motivating example throughout this paper, we consider the goal of estimating the effects of laws designed to reduce the lethality of opioid overdose. Specifically, we consider the research question: to what extent does enactment of a naloxone access law affect opioid overdose mortality rates in the United States (US)? (We define a naloxone access law as a law that allows naloxone, an opioid antagonist that reverses opioid overdose if administered in time,\cite{chamberlain1994comprehensive} to be prescribed and dispensed not only to people who use opioids, but also to laypersons such as nonmedical first responders, bystanders, and family/friends.\cite{davis2015legal})
 
Causal inference in policy evaluation can be challenging.\cite{goin2021guns,matthay2020revolution,schuler2020methodological,griffin2021methodological} To identify the (causal)\cite{hernan2018c} effects of a policy from observed data, one typically needs to invoke assumptions that include: exchangeability (meaning that there are no unobserved confounders of the policy exposure-outcome relationship), positivity (also called the common support condition\cite{abadie2018econometric} or overlap\cite{callaway2020difference}, defined in depth below), and no interference, (meaning that implementation of a policy exposure in one state cannot affect the counterfactual outcomes in other states),\citep{tchetgen2012causal} among others. All can be particularly difficult to satisfy in policy settings. Most policy evaluation research focuses primarily on addressing exchangeability through the use of difference-in-differences, synthetic control, and interrupted time-series methods, and some attention has been paid to interference in policy evaluation\cite[e.g.,][]{sobel2006randomized}. However, the threats to inference posed by positivity violations (further defined below) have been less discussed in epidemiologic health policy literature.\cite{matthay2020revolution,matthay2021everything}  Evaluating and addressing positivity violations is critical to reduce bias and improve inference in policy evaluation. 
 
In this paper, we focus on addressing challenges related to the positivity assumption.
\cite{petersen2012diagnosing} Consider a single timepoint setting and average treatment effect, denoted $\E(Y_1 - Y_0)$, where $\E$ denotes expectation and $Y_a$ denotes a counterfactual outcome under treatment value $A=a$. In this setting, positivity is defined as each unit having a positive probability of each exposure level used in the causal contrast (in this case, \{0,1\}) for every strata of observed confounder combinations, which can be denoted $P(A=a \mid W=w)>0$ for all $w$ for which $P(W=w)>0$, where $A$ denotes exposure, $W$ denotes covariates, and $P(A=a | W=w)$ denotes the probability of having exposure level $a$ conditional on observed covariates $w$ for all $(a,w)$ observed in the population of interest.
\cite{petersen2012diagnosing,hernan2020causal} In the context of our motivating example, positivity in a single timepoint setting would mean that every state has a nonzero probability of having a naloxone access law enacted 
(e.g., having ever enacted one) given other covariates, such as sociodemographic variables, the political climate, or enactment of other laws. The positivity assumption is inherently linked to the exchangeability assumption, because the covariates over which positivity must hold are the covariates needed to achieve conditional exchangeability.

The above defines structural positivity: assuming that the true probabilities are nonzero. However, practical violations of positivity (also called observed positivity violations) are also a concern, and occur when predicted probabilites of treatment conditional on covariates are close to, but not exactly, zero.\cite{petersen2012diagnosing} For example, in the above single timepoint setting, consider a state with covariates $w$ and exposure $A=0$. For positivity to hold practically, then at least one other state with the same covariates $w$ would need to have the other level of exposure, $A=1$ such that if one were to estimate the probability of exposure conditional on the covariates, the predicted probability would not be 0 or 1. Throughout this paper, our focus is on practical violations of the positivity assumption.

Because policies are typically enacted across states over time, a typical goal is to estimate policy effects in longitudinal settings; for example, the 
expected effect that having naloxone access laws in place over multiple years would have on opioid overdose rates in a later year. We could denote the expected effect on opioid overdose death rates in some later year, had such laws always been enacted everywhere vs. never been enacted anywhere, as $\E(Y_{\bar{1}} - Y_{\bar{0}})$, where $\bar{a}$ denotes a history of exposure $\bar{A}=\bar{a}$ over time, e.g., $\bar{a}_{t} = (a_0, a_1, a_2, ..., a_{t})$.  
For this longitudinal estimand, positivity is defined as each unit having a positive probability of each exposure level used in the causal contrast at each timepoint, $t$, for every strata of observed covariate history and exposure history in the causal contrast, which can be denoted $P(A_t=a_t | \bar{a}_{t-1}, \bar{l}_{t-1})>0$ for all $(\bar{a}_{t-1}, \bar{l}_{t-1})$ for which $P(\bar{A}_{t-1}=\bar{a}_{t-1}, \bar{L}_{t-1}=\bar{l}_{t-1})>0$.\cite{hernan2020causal} 
In terms of our motivating example, practical positivity in this longitudinal setting would mean that each state has a nonzero and not-too-small probability of 1) having a naloxone access law enacted in time $t$ given history (if available) of prior enactment and observed covariate history, including history of opioid overdose, and 2) not having a naloxone access law enacted in time $t$ given history (if available) of no prior enactment and observed covariate history. 
 For example, if a state has not yet enacted a naloxone access law, but has a history of rapidly increasing opioid overdose deaths and recently enacted other laws affecting overdose mortality rates, that state may have a close-to-zero predicted probability of not enacting a naloxone access law in that year. Thus, the example described would represent a practical violation of the positivity assumption. 

In this paper, we show empirical evidence of the practical positivity violations that can result when attempting to estimate effects of naloxone access laws. We discuss an alternative formulation to estimating policy effects that dramatically weakens the required positivity assumption, thereby offering a means to estimate policy effects even in scenarios with serious positivity problems. We also demonstrate that in many policy evaluations, the data structure is hierarchical, with the policy exposure occurring at a cluster level---e.g., at the state-level---and with covariates and outcomes measured at a finer-grained level---e.g., at the county-level. Respecting this structure by estimating a cluster- or state-level exposure model has important implications for practical positivity violations. 
 The paper is organized as follows. In the next section, we describe the motivating example of estimating the effect of enactment of naloxone access laws on subsequent opioid overdose rates. Then, we describe the importance of estimating the probability of policy implementation at the state-level on resulting practical positivity violations. Next, we provide an overview of positivity in the longitudinal policy evaluation setting, including evaluations of such violations in the context of our motivating example, and detail how such violations can be addressed. Last, we include analyses of our motivating example addressing positivity problems.

\section*{Motivating example}
\subsection*{Overview}
In response to the opioid overdose crisis\cite{hedegaard2017drug,wilson2020drug,hedegaarddrug,cdccovid} states adopted a range of laws, one of the most common being laws to increase access to naloxone.
Naloxone access laws typically include one or more of the following provisions: 1) permission for third-party prescriptions that allow healthcare practitioners and pharmacists to prescribe and dispense naloxone to people who are not at direct risk of overdose, 2) removal of liability for providing or administering naloxone, 3) permission for pharmacists to dispense naloxone to individuals without a patient-specific prescription, among other provisions.\cite{davis2015legal} Prior research has found mixed evidence for the effects of naloxone access laws on overdose mortality.\cite{doleac2019moral,erfanian2019impact,mcclellan2018opioid,abouk2019association,rees2019little,smart2021systematic} It is possible that one contributing factor to mixed findings could be violations of the identification assumptions, including positivity, particularly if other laws affecting overdose mortality were enacted close together in time with naloxone access laws.\cite{matthay2020revolution}

\subsection*{Data}
We use US county-level data 2007-2018 for all states and DC except Alaska (due to county boundary changes), and New Mexico and Connecticut (given their anomalous early enactment of naloxone access laws in 2001 and 2003, respectively). Thus, our sampling unit is county, although law enactment (and therefore, positivity) is at the state-level. We further describe data sources and details in Section e1 of the eAppendix.

\section*{Respecting the cluster-level policy exposure}
\label{sec:policylevel}
In our motivating example, and indeed in most epidemiologic research on the health effects of laws, the law is enacted at the state level (which we can denote $A^s$) and is a function of state-level covariates (denoted $A^s = f(W^s, U_A)$, where $U_A$ represents unobserved exogenous errors). Health outcomes are typically measured at a finer geographic level---e.g., county\cite{hamilton2021good,cerda2020measuring,cerda2021spatiotemporal}---and are a function of the state-level law being evaluated ($A^s)$, other state laws ($W^s$), as well as county-level covariates ($W$), denoted $Y = f(W, W^s, A^s, U_Y)$, where $f$ is some unknown function and $U_Y$ represents all other unmeasured causes of the outcome. 

As reviewed by Balzer et al., 2019,\cite{balzer2019new} one could conduct a state-level modeling approach with states as the units, and such an approach would be valid. However, it would not make full use of the county-level covariates, $W$, in estimating the outcome model, and thus, would result in an estimator with reduced power. Much of epidemiologic research on the health effects of laws\cite{hamilton2021good,cerda2020measuring,cerda2021spatiotemporal,doleac2019moral,blanchard2018state} adopts a county-level modeling approach with counties as the units. Under more restrictive assumptions than required for the state-level analysis, Balzer et al., 2019\cite{balzer2019new} showed that such a county-level analysis may also be a valid approach. However, the clustered nature of treatment assignment requires extra care in estimation of the treatment probabilities. Specifically, estimating the probability of law enactment at the county-level (using county-level variables as predictors of law enactment) results not only in a misspecified model, but can introduce practical positivity violations as an artifact of having many county-level units with identical exposures ($A^s$), clustered county-level covariates (values clustered within state), and identical state-level covariates ($W^s$). The state-level covariates in such a model are problematic, because the combination of values of $w^s$ perfectly or near perfectly identify the state, and thus, the state-level exposure. This issue is of particular relevance for inverse-probability of treatment weighted (IPTW) estimators and doubly robust estimators like augmented IPTW and targeted minimum loss-based estimators that make use of the estimated propensity scores. The correct strategy in this case is to estimate the state-level probabilities, using state level covariates in a state-level data set (i.e., estimate $P(A^s=1 | W^s)$). 
We believe that this potential problem is underappreciated. 

\begin{figure}
\caption{Practical positivity violations contrasting a (misspecified) county-level model with a state-level model.}
\label{fig:clusterlevel}
\subfloat[County-level model with state- and county-level covariates  \label{fig:clusterlevelcounty}]{
  \includegraphics[width=0.50\textwidth]{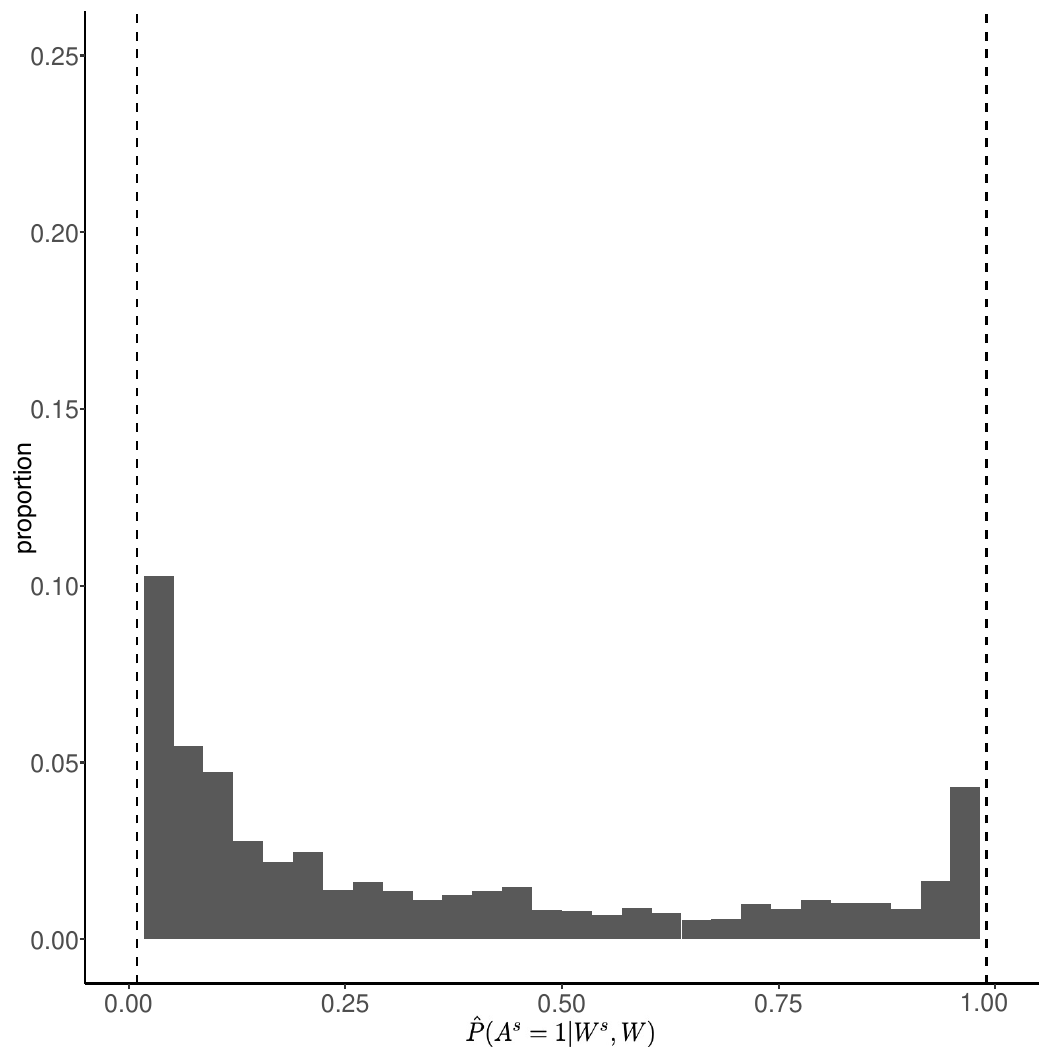}}
     \hfill
\subfloat[State-level model with state-level covariates \label{fig:clusterlevelstate}]{
\includegraphics[width=0.50\textwidth]{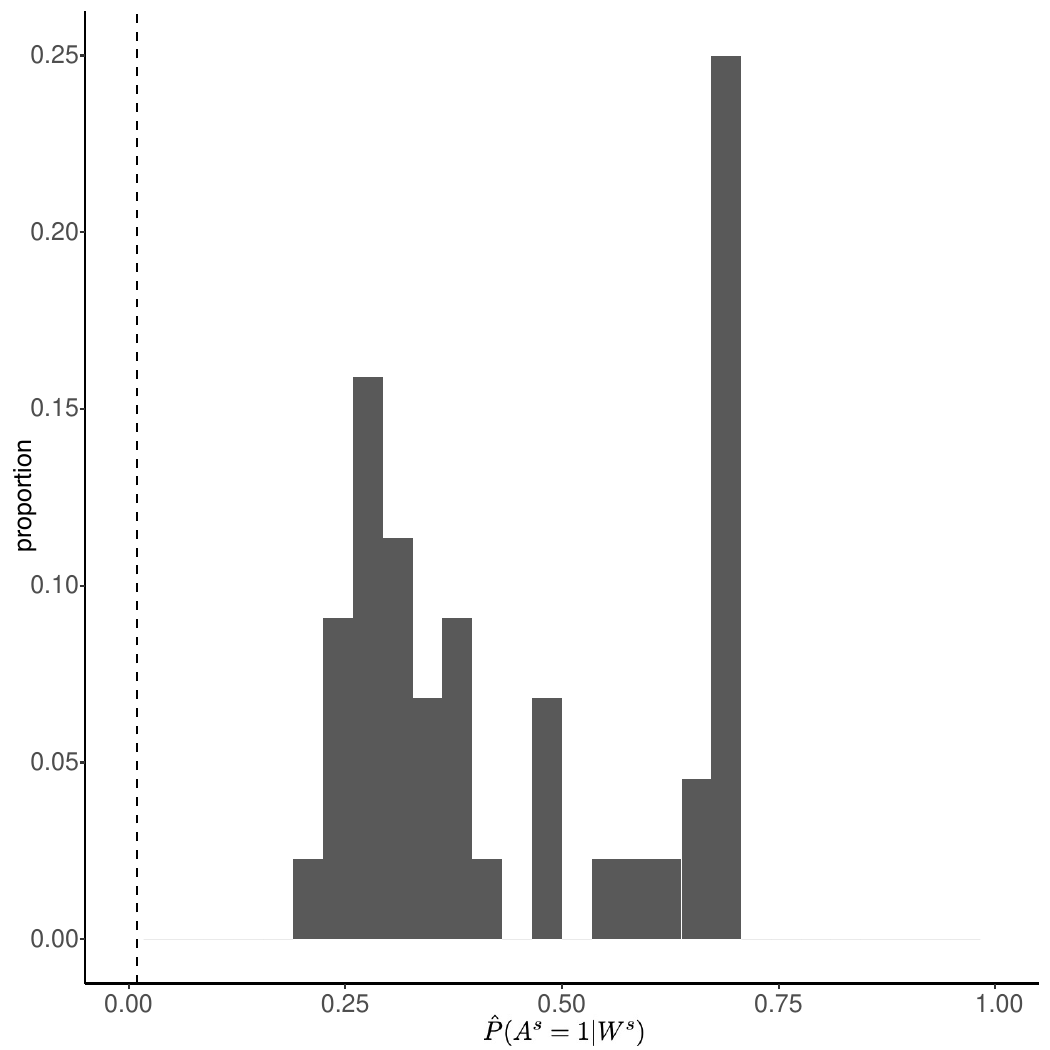}}\\
\subfloat[County-level model with county-level covariates only \label{fig:clusterlevelcountynostatecov}]{
\includegraphics[width=0.50\textwidth]{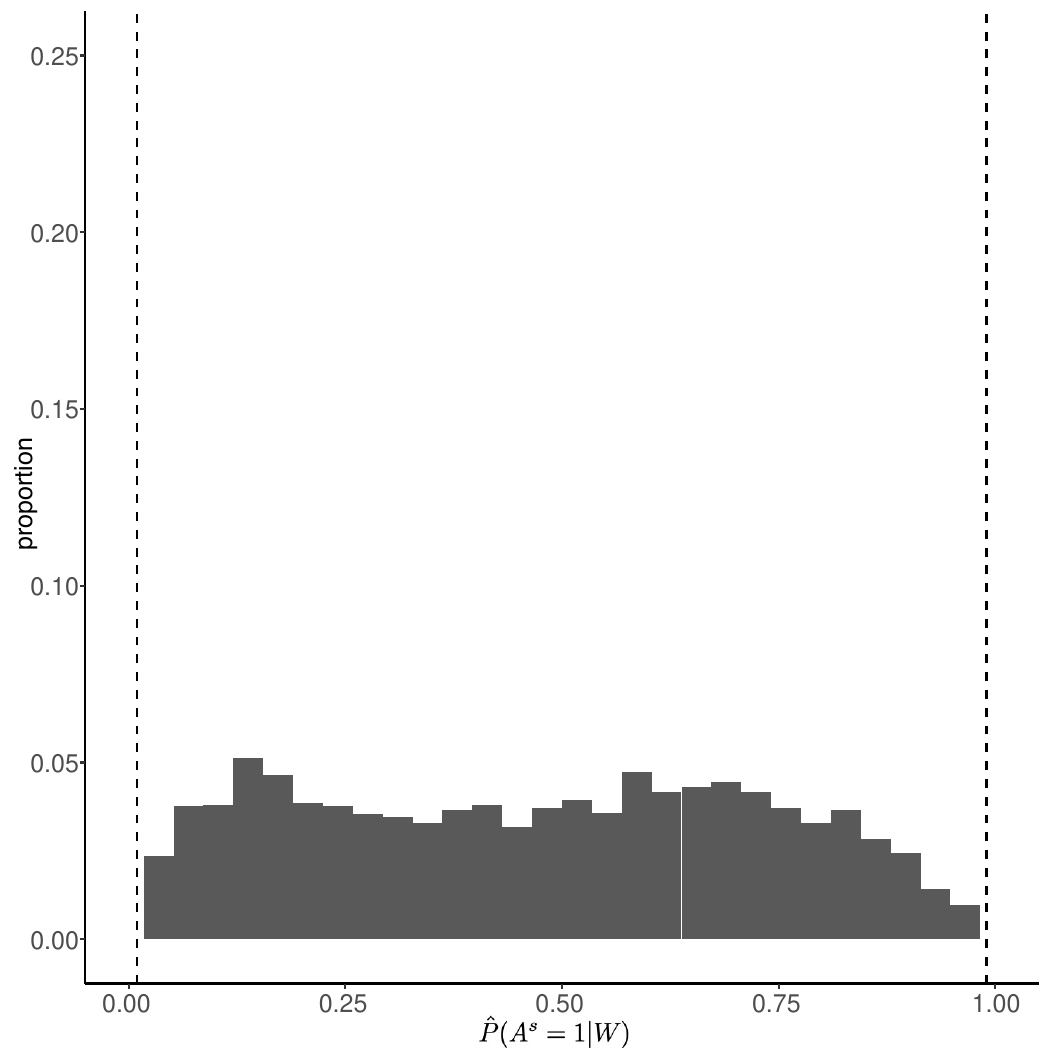}}
\end{figure}

We provide an example of how estimating the exposure model at the wrong level can induce practical positivity violations using our motivating example. For this example, we consider a simple, single timepoint setting. $A$ is a binary variable taking the value of 1 if the state had a naloxone access law enacted in or prior to 2014 and the value of 0 if the state did not have such a law enacted by 2014. Positivity in this setting is defined as a non-zero probability of having passed a naloxone access law in or prior to 2014 conditional on covariate combinations. Practical positivity means that this non-zero probability is also not too small. We can estimate these probabilities by fitting a model of this exposure measure conditional on covariate values and generating predicted probabilities. We do so both at the county-level (Figure \ref{fig:clusterlevelcounty}) and at the state-level (Figure \ref{fig:clusterlevelstate}). We use SuperLearner, which is an ensemble of machine learning algorithms (we included generalized linear models, lasso, and multiple additive regression splines) that weights each algorithm to result in the lowest mean prediction error.\cite{van2007super} 

We see dramatic evidence of practical positivity violations when the propensity for law enactment was estimated at the county level. Fully 33\% of counties have predicted probabilities <0.01 or >0.99 of having enacted naloxone access laws in 2014 or earlier given their covariate values. In contrast, 
in the state-level model, none of the states have predicted probabilities <0.01 or >0.99. We can also see in Figure \ref{fig:clusterlevelcountynostatecov} that the practical positivity violations in the county-level model are an artifact of the state-level covariates, because when we remove these covariates, 
only 0.1\% of the counties have predicted probabilities $<0.01$ or $>0.99$.
 
Although using a county-level model of the law exposure and excluding the law covariates would mostly resolve the artifactual violations of the positivity assumption (Figure \ref{fig:clusterlevelcountynostatecov}, such a strategy would be problematic for at least two reasons. First, exchangeability would no longer hold because the inclusion of the state-level law covariates is needed to satisfy the conditional exchangeability assumption.\cite{abadie2018econometric} Without these covariates, the effect of naloxone access law enactment would be conflated with enactment of the excluded laws. 
For example, Good Samaritan laws, which provide some legal protections to those witnessing or experiencing an overdose to encourage help-seeking,\cite{PDAPS,davis2015legal} are frequently enacted in conjunction with a naloxone access law (in 31\% of all states in the US) or within 2 years of naloxone access law enactment (in 63\% of states). If we did not control for Good Samaritan laws, we would not know which portion of the estimate was due to naloxone access laws and which portion was due to Good Samaritan laws. Others have demonstrated that this strategy, though common in the policy literature,\cite[e.g.,][]{mcclellan2018opioid, bao2016prescription,maughan2015prescription} can result in large biases.\cite{griffin2021methodological} The other reason it would be problematic is because the model would be misspecified. The law exposure is a state-level variable, and is a function of state-level predictors. 

\section*{Positivity in the longitudinal setting}
\label{sec:long}
Many policy evaluations take advantage of longitudinal data and then exploit point-in-time policy enactment as an exogenous or conditionally exogenous interruption around which differences (interrupted time series) or difference-in-differences can be estimated.\cite{abadie2018econometric,benmarhnia2019rose,abadie2010synthetic,doudchenko2016balancing} Typically, such approaches are used to answer research questions that ask what happens to a particular unit (e.g., state) or group of units that implemented a policy vs. what would have happened if the policy had not been implemented, using non-implementing units as controls. In discussing how this relates to our applied example, we introduce some notation. We can assume our observed longitudinal naloxone access law data is as follows: $O=(L_1, A_1, L_2, ..., A_T, Y),$ where $L_1$ represents baseline covariates, including the baseline measure of the outcome; $A_t$ represents a binary 0/1 exposure of whether or not the state had a naloxone access law in effect in year $t$; $L_t$ represents time-varying covariates at time $t$, including the outcome measure at time $t-1$; and $Y$ is the outcome at the final year, $T$.

\begin{figure}
\caption{Naloxone access law enactment by state (Alaska, Connecticut, New Mexico excluded).}
\label{fig:enact}
\centering
  \includegraphics[width=0.75\textwidth]{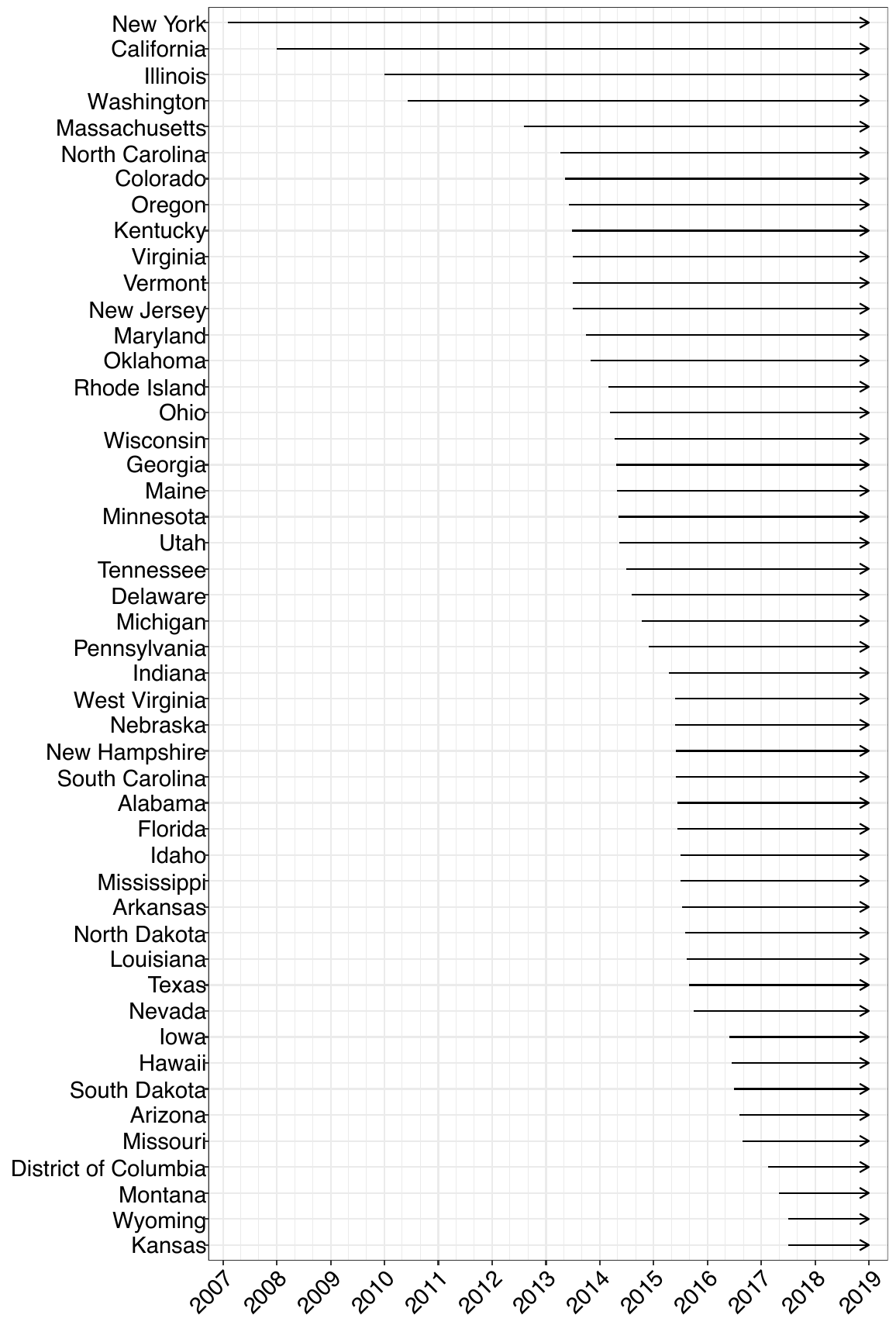}
\end{figure}

In the case of naloxone access laws, every state in the US (and DC) enacted a naloxone access law provision by 2017. 
 Figure \ref{fig:enact} shows naloxone access law enactment by state over time. Using analytic methods typical of policy analysis, one could envision a serial difference-in-differences-type analysis (another example is synthetic control and its variations) where short-term effects are estimated for each state or group of states at each time of enactment, choosing a set of control states that did not implement at any of the years in question. See \cite{goodman2021difference,callaway2020difference} for examples. However, the group of possible control states shrinks over time and for the last six states to implement, no control states exist. A common alternative is to use a two-way fixed effects analysis with fixed effects for each unit (e.g., state) and time (e.g., year). However, such an approach makes a parallel trends assumption---meaning that the relationship of time with the expected outcome in the absence of the policy is constant across units---unless the state fixed effects interact with year fixed effects, at the price of potentially severe data sparsity issues.\cite{abadie2018econometric} 

\begin{figure}
\caption{Paths estimated using longitudinal estimands that account for time-varying treatment-confounder feedback versus using panel data-type estimands that do not.}
\label{fig:seqvpanel}
\subfloat[Paths estimated accounting for time-varying treatment-confounder feedback]{
  \includegraphics[width=0.50\textwidth]{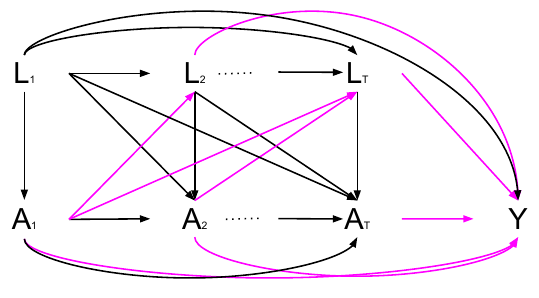}}
     \hfill
\subfloat[Paths estimated not accounting for treatment-confounder feedback]{
\includegraphics[width=0.50\textwidth]{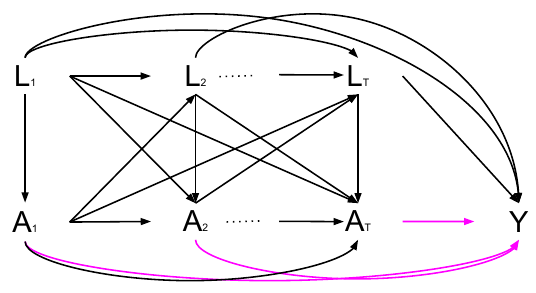}}\\
\end{figure}

So-called g-methods for longitudinal effect estimation\cite{robins1986new} are a more common approach in epidemiology and represent an alternative to the above econometric methods. Such methods include longitudinal g-computation (also called the longitudinal g-formula), IPTW estimators, and longitudinal double robust estimators, like augmented-IPTW and targeted minimum loss-based estimators. The main appeal of such methods is that they account for dynamic feedback between time-varying exposures and time-varying confounding variables (including intermediate measures of the outcome), in which confounders may affect exposure, and exposure may affect confounders over time.\cite{hernan2020causal} These methods 
generally rely on 
 sequential regression to correctly adjust for time-varying confounders affected by prior exposure\cite{robins1986new,bang2005doubly,van2012targeted}---that is, previous confounders, $\bar L_{t}=(L_1,\ldots,L_t),$ are adjusted for in estimating the relationship between a particular $A_t$ and $Y$, but the effect of $A_t$ on $Y$ is allowed to operate through subsequent time-varying confounders, $(L_{t+1}, \ldots,L_T)$, that are on the causal pathway from $A_t$ to $Y$. Figure \ref{fig:seqvpanel} contrasts longitudinal, path-specific effects estimated by panel-data-type methods with longitudinal effects estimated by g-methods.

A typical longitudinal estimand in the epidemiologic causal inference literature is the longitudinal average treatment effect (ATE) contrasting an always-treat versus never-treat treatment strategy,\cite{hernan2020causal,van2011targeted} which can be denoted $\E(Y_{\bar{1}} - Y_{\bar{0}})$, using $Y_{\bar{a}}$ to denote the counterfactual outcome had the exposure history over time (denoted by $\bar{A}$) been equal to $\bar{a}$, contrasting $\bar{1}$ with $\bar{0}$.  We focus on such a causal effect here, but note that other longitudinal causal estimands are typical in the econometrics literature. For example, the popular two-way fixed effects difference-in-differences model reflecting staggered adoption of a policy over time estimates a weighted average treatment effect on the treated (ATT) estimand, under certain assumptions.\cite{goodman2021difference} This approach does not include the portion of the exposure effect that operates through subsequent time-varying confounders.\cite{hernan2020causal} 

In the case of our motivating example, the always-treat vs. never-treat ATE ($\E(Y_{\bar{1}} - Y_{\bar{0}})$) would involve contrasting the counterfactual outcome had naloxone access laws been enacted in all 11 years prior to 2018 versus if it had not been enacted in any of the previous 11 years: $\bar{1}=(1,1,1,1,1,1,1,1,1,1,1)$ with $\bar{0}=(0,0,0,0,0,0,0,0,0,0,0)$. To appropriately account for time-varying confounding, one would need to rely on the sequential exchangeability assumption (denoted $Y \indep A_t \mid \bar{A}_{t-1}=\bar{a}_{t-1}, \bar{L}_t$), and positivity (denoted $P(A_t=a_t | \bar{a}_{t-1}, \bar{l}_{t-1})>0$, discussed in the Introduction). In words, assuming an absence of practical positivity violations in this scenario would mean assuming that there would be a nonzero and not-too-small probability of 1) having a naloxone access law enacted at each timepoint given history, if it exists, of enactment and past observed covariate history ($P(A_t=1 | \bar{a}_{t-1} = 1, \bar{l}_{t-1})>0$) and 2) not having such a law enacted at each timepoint given history, if it exists, of no enactment and past observed covariate history ($P(A_t=0 | \bar{a}_{t-1} = 0, \bar{l}_{t-1})>0$).\cite{hernan2020causal}  

\begin{figure}
\caption{Practical positivity violations in a longitudinal state-level model.}
\label{fig:poslongdet}
\subfloat[Practical positivity violations for every state having naloxone access laws enacted all years ($\bar{1}$) \label{fig:poslongdet1}]{
  \includegraphics[width=0.50\textwidth]{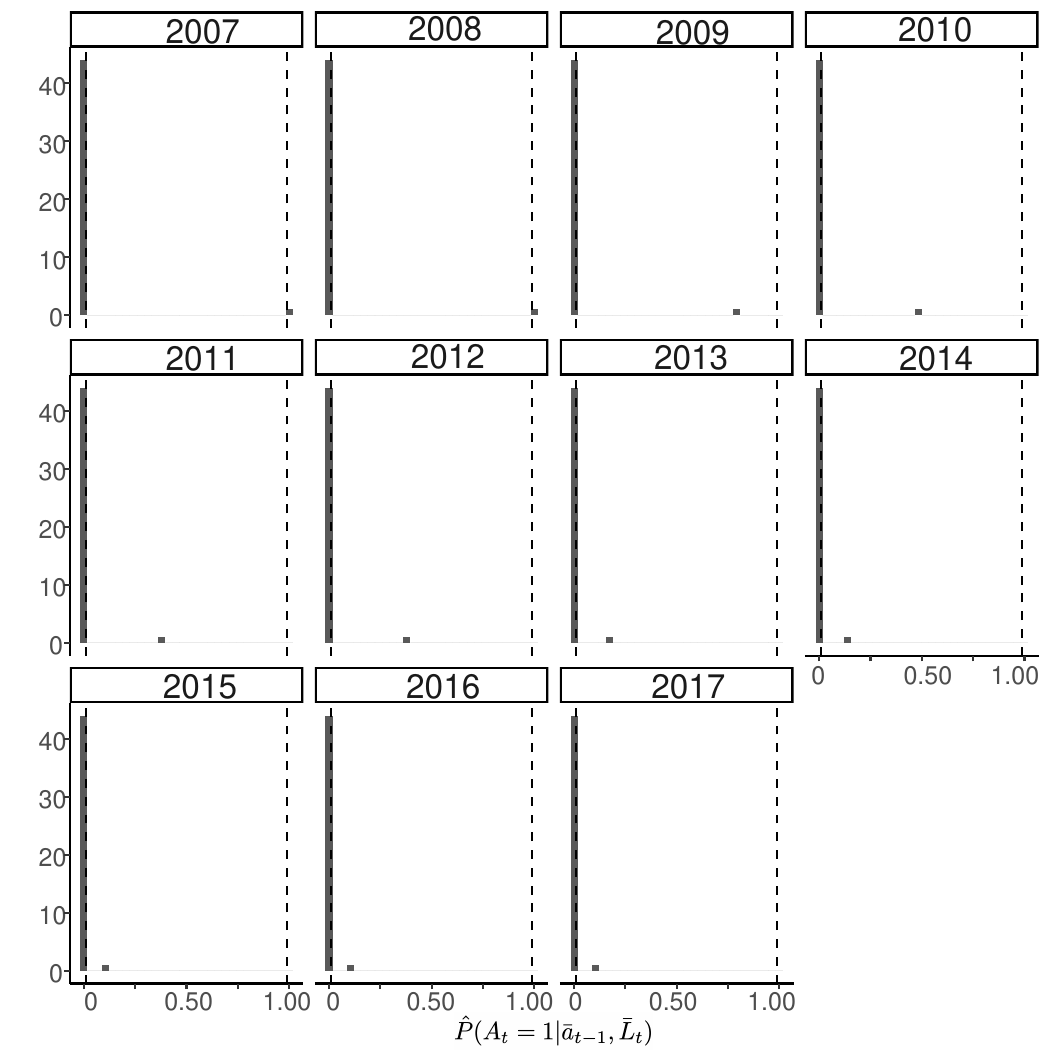}}
     \hfill
\subfloat[Practical positivity violations for no state ever having a naloxone access law enacted in any year ($\bar{0}$) \label{fig:poslongdet0}]{
\includegraphics[width=0.50\textwidth]{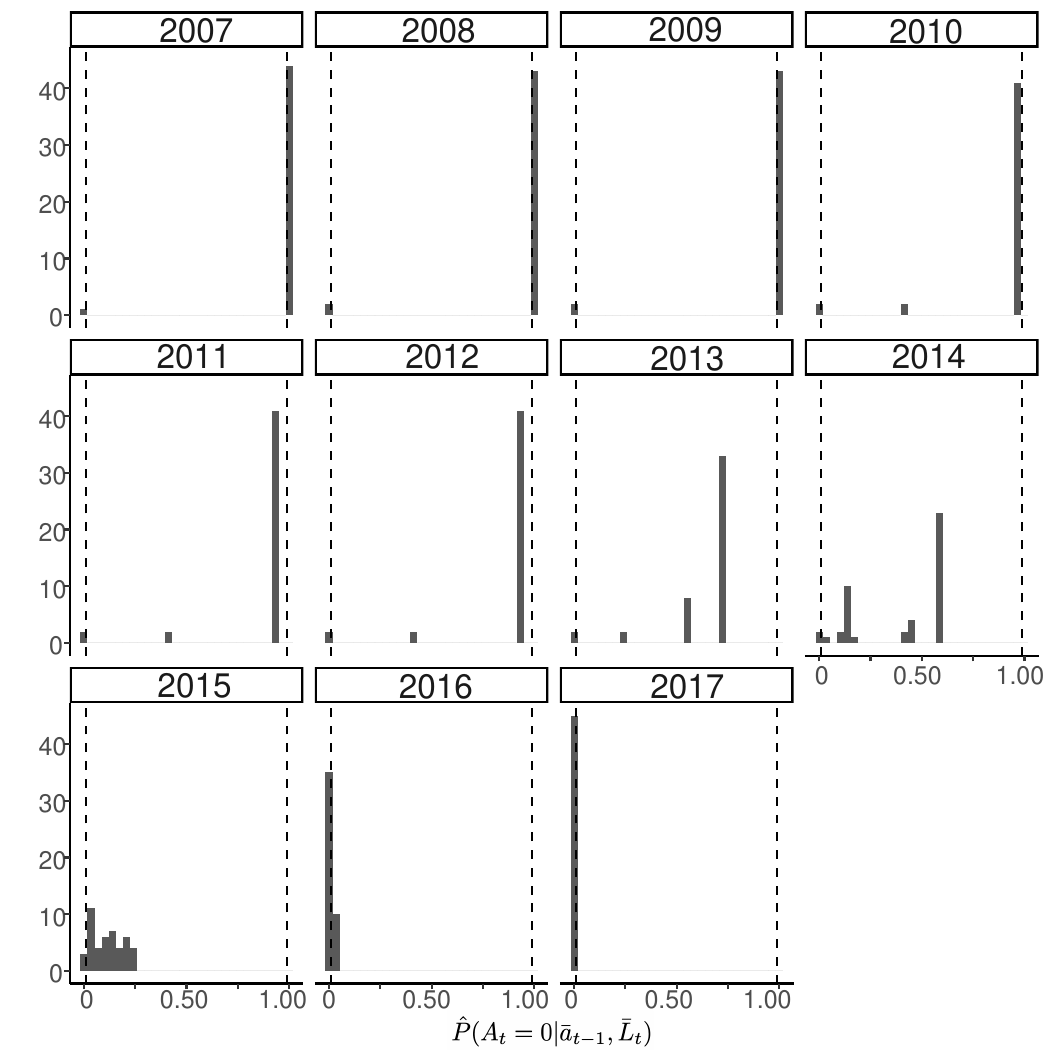}}\\
\end{figure}

Using an appropriate state-level model, 
we see in Figure \ref{fig:poslongdet} dramatic evidence of practical positivity violations. For the $\bar{1}$ portion of the contrast, we see in Figure \ref{fig:poslongdet1} that all states except one have very small predicted probabilities of enacting a naloxone access law in the first year of 2007. These low probabilities are compounded over time. There is better support for the $\bar{0}$ portion of the contrast. We see in Figure \ref{fig:poslongdet0} that only one state has a predicted probability <0.01 of having a naloxone access law enacted in years 2007-2015. However, in 2016, over 40\% of states have predicted probabilities <0.01, and in 2017, all states have predicted probabilities <0.01. We note that the positivity assumptions necessary to identify a weighted ATT estimand from a staggered adoption difference-in-differences setting would differ. Namely in such a setting, for all observed $t$, the probability of the law first being implemented at time $t$ conditional on pre-implementation covariates and on not having the law implemented prior to time $t$ must be bounded away from 1.\cite{callaway2020difference} 

Frequently, researchers ignore positivity violations. 
However, doing so is not advisable 
because it would mean 
estimating an effect that is not identified from the observed data. 
 If estimation proceeds anyway, 
 the resulting estimates can be substantially biased. 
 For estimators that are based on the outcome model, practical positivity violations can induce bias due to extrapolation if there are units in one exposure group without comparable counterparts in the other exposure group---also called a lack of overlap. In this case, predicting the counterfactual outcome for the units without the comparable counterparts would rely on extrapolating outside of the region over which the model was fit. If the true model in the extrapolation region differs from the fitted model, then bias may result.
For estimators that are based on the treatment or exposure model (e.g, IPTW estimators and doubly robust estimators), positivity violations induce finite sample bias in cases where very few units in one exposure group serve as the comparable counterpart for many units in the other exposure group. Relying on few units for estimation can result in estimates that are highly variable, especially for estimators that use inverse probability weighting, meaning that standard errors are very large, sometimes even rendering the estimate no longer informative, due to the low statistical power.\cite{robins2007comment,petersen2012diagnosing} 

In the face of such widespread practical positivity violations, an alternative strategy is to reformulate the estimand from one that explicitly compares states, as the above ATEs would entail, to instead correspond to the effect of shifting when the naloxone access law was enacted in each state, corresponding to what is called a ``modified treatment policy''.\cite{Haneuse2013,diaz2020non}
We can define an alternative longitudinal causal effect that corresponds to such a longitudinal modified treatment policy\cite{diaz2020non} 
by considering an estimand that compares the counterfactual opioid overdose mortality rate had states delayed their enactment of naloxone access laws by 1 year versus what was observed. Redefining the estimand in terms of a 1-year exposure delay may be able to identify the longitudinal effect of naloxone access law enactment on opioid overdose mortality. Specifically, for this causal estimand, we are interested in the effect on county-level opioid overdose mortality rate in 2018 ($Y$) had states enacted their naloxone access laws 1 year later, adjusting for time-varying state-level covariates in the exposure models and adjusting for state-level and county-level covariates in the outcome models. The estimand corresponding to this longitudinal modified treatment policty can be written $E(Y_{\bar{\dd}}) - E(Y)$, where $\dd_t$ is defined as: 
 \begin{equation}\label{eq:defdshiftlong}
    \dd_t(a_t, a_{t-1}) =
    \begin{cases}
      a_t = 0 & \text{if } a_{t-1}=0 \text{ and } a_t=1, \\
      a_t            & \text{otherwise}, 
    \end{cases}
  \end{equation}
for all $a_t \in \{0,1\}$ and all $t \in \{2007-2017\}$. To summarize, this reformulation moves away from an always vs. never ($\bar{1}$ vs. $\bar{0}$) comparison and instead compares hypothetical outcomes under a shift in exposures---in this case, delaying a yearly binary law exposure by 1 year, or in other words, shortening the duration by 1 year. (Note: we estimate the effect of delaying the law by one year instead of implementing it one year earlier for technical reasons related to identification.) 

\begin{figure}
\caption{Positivity in the longitudinal modified treatment policy estimand (stochastic counterfactual outcomes) setting. Density ratio is defined as $\frac{g_\delta(A_t \mid \bar{A}_{t-1}, \bar{L}_t)}{g(A_t \mid \bar{A}_{t-1}, \bar{L}_t)}.$}
\label{fig:posnallong}
\centering
\includegraphics[width=.5\textwidth]{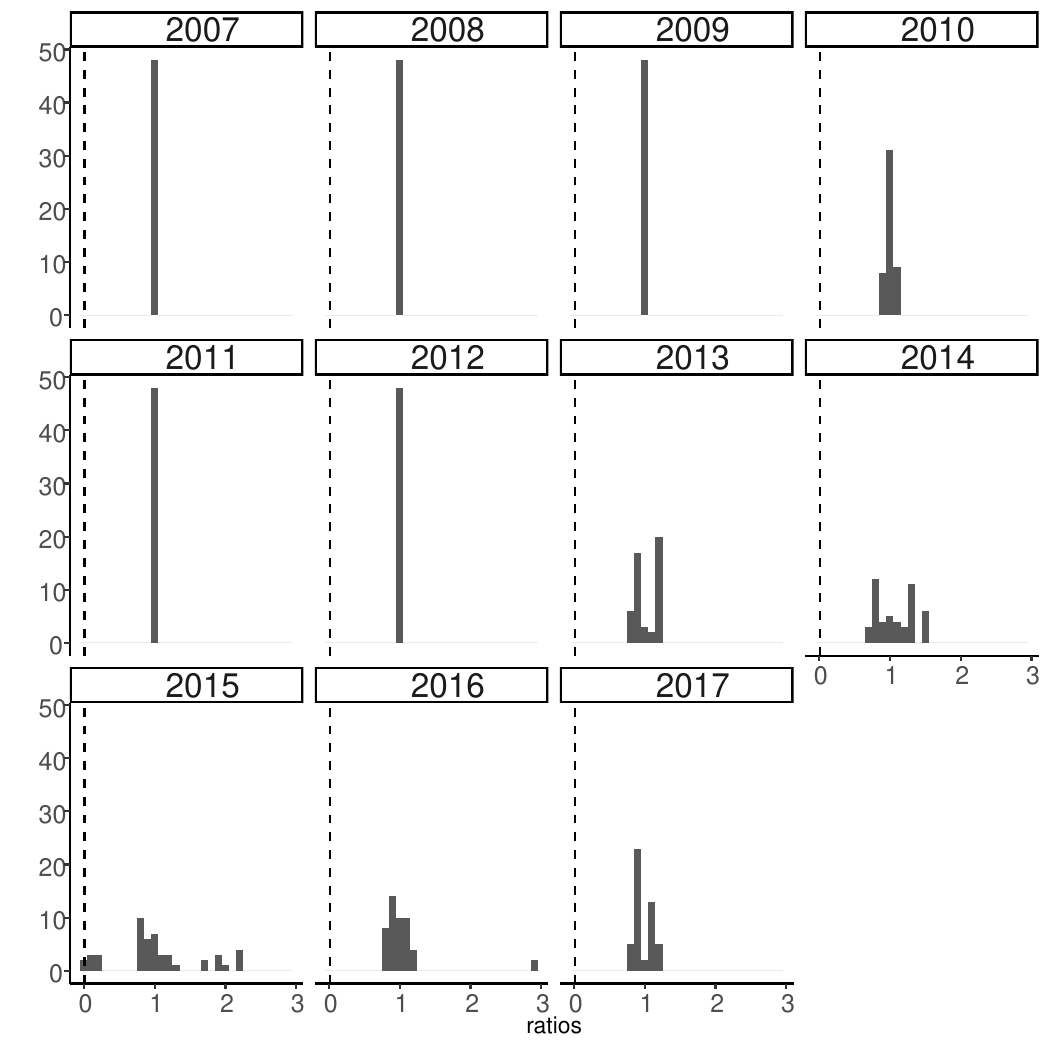}
\end{figure}

Consequently, this longitudinal effect of shifting the year of enactment of a naloxone access law weakens the positivity assumption from relying on the stability of $\prod^T_{t=1} \frac{I(A_t=a_t)}{g(A_t=a_t \mid \bar{a}_{t-1}, \bar{l}_t)}$ for $a_t \in \{0,1\}$ to instead rely on the less variable ratio: $\prod^T_{t=1}\frac{g_\delta(A_t \mid \bar{A}_{t-1}, \bar{L}_t)}{g(A_t \mid \bar{A}_{t-1}, \bar{L}_t)}$, where $g_\delta(A_t \mid \bar{A}_{t-1}, \bar{L}_t)$ is the density function of the random variable $A_{t,\delta} = \dd_t(A_t, A_{t-1})$ (see D\'iaz et al., 2020\cite{diaz2020non} for additional description and details). The predicted cumulative density ratios at each year are shown in Figure \ref{fig:posnallong}. Figure \ref{fig:posnallong} shows that there is no evidence of practical positivity violations in any of the years. 


\section*{Data Analysis}
We now estimate the longitudinal modified treatment policy estimand introduced in the previous section: the effect of having delayed enactment of a naloxone access law by one year on opioid overdose mortality rates in 2018, accounting for time-varying confounding, denoted $\E(Y_{\bar{\dd}}) - \E(Y)$. Of note, 31\% of states enacted a Good Samaritan law within the same year as enactment of a naloxone access law. 
This effect is identified under the following assumptions: 1) sequential exchangeability, defined in the previous section; 
 2) positivity---that at each time $t$ the ratio $\prod^T_{t=1}\frac{g_\delta(A_t \mid \bar{A}_{t-1}, \bar{L}_t)}{g(A_t \mid \bar{A}_{t-1}, \bar{L}_t)}$ is defined; and 3) at each time $t$, no county's counterfactual outcome is affected by another state's exposure history, which is the no interference assumption. 
 
 Assuming exchangeability may be reasonable in this case, as we have included numerous sociodemographic summaries at the county level and other related policy variables at the state level, consistent with the majority of the related literature.\cite{castillo2019prescription, hamilton2021good, blanchard2018state} However, unobserved confounding may nonetheless persist; for example, we are not accounting for political climate. We discussed the positivity assumption in this case in the previous section. In regards to the interference assumption, all counties within the same state have the same exposure values. However, interference is likely present in that it is plausible that the counterfactual opioid overdose mortality rates in some counties could be affected by the presence of policies in neighboring states.\cite{erfanian2019impact} It is unclear the extent to which this would bias estimates. The no interference assumption of no interference has been ignored in the vast majority of this literature,\cite{doleac2019moral,mcclellan2018opioid,blanchard2018state,abouk2019association,rees2019little,smart2021systematic} possibly in part because there are few practical solutions for addressing it. Most solutions proposed involve assuming some independent units.\cite{ogburn2017causal} 
 We could group states into regions, but at the cost of greatly reduced sample size and the unlikely plausibility of independent regions.

To estimate the additive effect contrasting this longitudinal modified treatment policy  (delay enactment by 1 year) with what was observed, we use a cross-fitted (5 folds), 
sequentially doubly robust estimator,\cite{diaz2020non} again including an ensemble of machine learning algorithms (generalized linear models, lasso, and multiple additive regression splines). Standard errors and confidence intervals were calculated using the sample variance of the efficient influence curve, modified to reflect the clustering of counties within states. Although this estimator was developed previously,\cite{diaz2020non} we describe it in Section e2 of the eAppendix. We also provide commented code that implements this estimator in our data: \url{https://github.com/kararudolph/code-for-papers/blob/master/NALeffects.R}.

 We estimate that delaying enactment of a naloxone access law by 1 year is associated with a slight increase in opioid overdose mortality rate in 2018 of 1.51 more opioid overdose deaths per 100,000 individuals aged 12 and over (95\% CI: -0.159, 3.18).
 
\section*{Discussion}
In this article, we described how positivity violations threaten the identification and estimation of policy effects. Using our motivating example of estimating the effect of naloxone access laws on opioid overdose mortality rates, we assessed and quantified threats to the positivity assumption. Such threats were sometimes dramatic---for example, in the always-treat portion of the longitudinal effect contrast, all US states except one had practical positivity violations. We also discussed and demonstrated how 
using a county-level model for a state-level exposure could result in artifactual practical positivity violations.

After assessing and establishing that practical positivity violations were an issue in the longitudinal estimation of the effect of enactment of a naloxone access law on later opioid overdose mortality rates, the next question was what to do about it. We offered two suggestions. 

The first suggestion was to model the exposure mechanism at the appropriate cluster level. Enactment of a naloxone access law is a state-level exposure influenced by state-level covariates. Modeling it as such was important to avoid inducing practical positivity violations.

The second suggestion was to redefine the estimand to correspond to a shift intervention or a modified treatment policy. Modified treatment policy estimands are defined and estimator options exist for both the single and multiple timepoint settings.\cite{Haneuse2013,munoz2012population,diaz2020non} However, to our knowledge, they have not been widely used in epidemiologic studies. This presents an opportunity, as these estimators dramatically weaken the positivity assumption to offer a way forward if practical violations of the positivity assumption are extensive. In the longitudinal setting, such longitudinal modified treatment policies and their estimation approaches also appropriately account for time-varying confounders and the dynamic feedback that results.\cite{diaz2020non,van2012targeted,hernan2020causal}

We applied both strategies to estimate the longitudinal effect of delaying enactment of a naloxone access law by 1 year on opioid overdose rates in 2018. Our results suggest that delaying enactment by 1 year would be associated with 1.51 more opioid overdose deaths per 100,000 individuals aged 12 and over (95\% CI: -0.159, 3.18), thus providing some evidence in favor of naloxone access laws' effectiveness on reducing opioid overdose mortality rates. Previous evaluations of the effectiveness of these laws 
 demonstrated mixed results in terms of whether or not they increased,\cite{erfanian2019impact} decreased,\cite{mcclellan2018opioid,blanchard2018state,abouk2019association,rees2019little,cataife2019regional} or had no overall effect\cite{cataife2019regional,doleac2019moral} on overdose mortality rates.\cite{smart2021systematic} 
 All but one of the previously mentioned studies controlled for co-occurring state-level laws.\cite{doleac2019moral,erfanian2019impact,blanchard2018state,abouk2019association,rees2019little} The remaining study chose not to control for co-occurring laws, conflating their potential effects with those of naloxone access laws.\cite{mcclellan2018opioid} 

However, our effect estimates are intended for tutorial purposes and should not be interpreted causally, as they remain plagued by other threats to validity like unobserved confounding and interference. It is unlikely we measured all of the relevant drivers of the opioid overdose epidemic. 
 For example, we did not control for any broader social/economic laws, even though these could be related to how long a naloxone access law or Good Samaritan law had been in place and related to opioid overdose rates. Future work could focus on better addressing endogeneity, perhaps by focusing on lethality of opioid overdose as the outcome (proportion of opioid overdoses that are fatal), attempting to control for opioid supply factors such as fentanyl penetration into the opioid market. However, data for nonfatal overdoses and for fentanyl supply can suffer from measurement error issues.\cite{eisenberg2019use,rowe2017performance,gilbert2017silicon,slavova2019methodological} 
 In addition, 
 interference is likely,\cite{erfanian2019impact} but has been ignored in the vast majority of this literature.\cite{doleac2019moral,mcclellan2018opioid,blanchard2018state,abouk2019association,rees2019little,smart2021systematic} We discussed interference in terms of our motivating example in the previous section. Briefly, the naloxone access law exposure is at the state level, and interference between states is likely, but without a good solution, and the resulting bias is unknown. 

Estimating the health effects of policies is a practical, important endeavor. It also entails multiple challenges to causal inference. Positivity violations, which may be in part due to policy co-occurrence, have been frequently ignored in previous epidemiologic health policy research. 
These issues have been highlighted in some recent epidemiologic\cite{matthay2020revolution,schuler2020methodological} and biostatistical\cite{diaz2020non} publications. We build on this recent work and demonstrate how positivity violations can be assessed and addressed in policy evaluation. Inference, however, 
remains a critical hurdle and area for future work.

\bibliographystyle{plain}
\bibliography{references}

\newpage
\appendix

\section{Data Details and Sources}
\subsection{Exposure: NAL} 
The Prescription Drug Abuse Policy System\cite{PDAPS} provided state data on effective dates of laws, including NAL provisions. 
We operationalized NAL as a binary 0/1 variable indicating whether any of the four provisions was in effect for the majority of the calendar year.
        
\subsection{Outcome: opioid overdose mortality rate} 
Opioid overdose deaths per year were obtained from restricted cause of death files from the National Center for Health Statistics,\cite{NCHS} and included the following ICD-10 codes\cite{ICD10}: X40-44, X60-64, X85, Y10-14, T40.0-T40.4, and T40.6. The mortality rate was the number of deaths per 100,000 of the county population aged $\ge12$ years, obtained from the National Center for Health Statistics.\cite{NCHS}   

\subsection{Covariates} 
\label{sec:cov}
County-level covariates included: the proportion of families under the poverty threshold; proportion unemployed; median household income; 
proportion male; proportion white, black, Hispanic/Latino; population density; proportion in the following age categories: 0-19, 20-24, 25-44, 45-64, 65+, all obtained from GeoLytics.\cite{geolytics} We included the outcome measured at baseline as a covariate as well. At the state-level, covariates included laws that could be associated with NAL enactment and with overdose mortality: enactment of any Good Samaritan law (GSL),\cite{PDAPSGSL} 
pain management clinic law (PMCL),
\cite{PDAPSPMC} medical marijuana law (MML) (allowing home cultivation or operational dispensary),
\cite{PDAPSMML,MPP,pacula2015assessing} 
and prescription drug monitoring law (PDMP),
\cite{horwitz2020importance} where the data sources are given in the citations. GSL, PMCL, and MML were operationalized as the proportion of the year the law was enacted, $[0,1]$, and PDMP ranged from 0 - 2, summing proportion of year a state had a modern operational PDMP system and proportion of year a state had a law that required PDMP system querying prior to prescribing or dispensing. 

\section{Estimator}

To illustrate the estimation strategy employed in this paper, we focus on a simple example with only two time points. We start by describing a sequential regression estimator, and then extend it to describe the sequentially doubly robust estimator used in this paper. The sequential double robustness means that, at each timepoint, if the exposure or outcome model is correctly specified, then the estimator is consistent (meaning the estimate would be expected to be unbiased).\cite{luedtke2017sequential,rotnitzky2017multiply} 

Consider data $L_0, A_1, L_1, A_2, Y$, where $Y$ is the outcome of interest. We will use $H_1=L_0$ and $H_2=(L_0,A_1, L_1)$ to denote the history of the data up to right before $A_1$ and $A_2$, respectively. Longitudinal modified treatment policies assess the effect of hypothetical interventions in which the exposure would have been assigned in terms of a user-given function $\dd_t(A_t, H_t)$. In our example, we use
\begin{equation*}
    \dd_t(a_t, a_{t-1}) =
    \begin{cases}
      a_t = 0 & \text{if } a_{t-1}=0 \text{ and } a_t=1, \\
      a_t            & \text{otherwise}. 
    \end{cases}
\end{equation*}
This function assigns exposure $A_t=0$ for units that were treated at time $t$ ($A_t=1$) and were previously untreated ($A_{t-1}=0$). For units that were untreated at time $t$ ($A_t=0$) or were previously treated ($A_{t-1}=1$), this function does not make any change to treatment, i.e., it allows $A_t$ to take its natural value. 

A sequential regression estimator for this parameter would proceed in the following steps:
\begin{enumerate}
    \item Fit a regression model $\hat m_2(a_2, h_2)=\hat E(Y\mid A_2=a_2, H_2=h_2)$. That is, regress the outcome $Y$ on  variables $A_2$ and $H_2$.
    \item Predict the outcome under the hypothetical intervention. That is, compute the prediction $\hat Y_{2,\dd}=\hat m_2(\dd_2(A_2, A_1), H_2)$ for every unit in the sample.
    \item Fit a regression model $\hat m_1(a_1, h_1)=\hat E(\hat Y_{2,\dd}\mid A_1=a_1, H_1=h_1)$. That is, regress the pseudo-outcome $\hat Y_{2,\dd}$ on  variables $A_1$ and $H_1$.
        \item Predict the outcome under the hypothetical intervention. That is, compute the prediction $\hat Y_{1,\dd}=\hat m_1(\dd_1(A_1, A_0), H_1)$ for every unit in the sample.
        \item  The final estimate of the mean outcome under the hypothetical intervention is the average over the entire sample of $\hat Y_{1,\dd}$.
\end{enumerate}

The above sequential regression estimator works well if we are able to a-priori and correctly specify a parametric model (e.g., logistic regression) for the regressions in steps 1 and 3, but it can be biased if these models are wrong. 

To mitigate model misspecification bias, \cite{diaz2020non} developed a \textit{sequentially doubly robust estimator} that has the following properties:
\begin{itemize}
    \item It incorporates the exposure mechanism $g_t(a_t\mid h_t)=P(A_t=a_t\mid H_t=h_t)$ into the estimation procedure,
    \item It is doubly robust in the sense that it remains consistent (i.e., approximately unbiased in large samples) if, at each time point, either $m_t$ or $g_t$ is correctly estimated,
    \item It can incorporate flexible regression (e.g., machine learning, model selection techniques, splines, etc.) into the estimation procedure to improve the chances of correct model specification.  
\end{itemize}

To introduce the estimation procedure, it is necessary to introduce the concept of a \textit{sequentially doubly robust transformation}. For $t=2$, a sequentially doubly robust transformation $\psi_2(Z)$ is a function of the data (where we use $Z$ to denote the data) and $(m_2, g_2)$ that satisfies:
\[m_1(a_1, h_1) = E[\psi_2(Z)\mid A_1=a_1, H_1=h_1]\]
whenever either $m_2$ or $g_2$ are the correct models. For time $t=2$ the doubly robust transformation is equal to:
\[\psi_2(Z)=\frac{g_2(\dd(A_2,A_1), H_2)}{g_2(A_2, H_2)}(Y-m_2(A_2, H_2)) + m_2(\dd(A_2,A_1), H_2).\]
The corresponding formula for $\psi_1(Z)$ is given in \cite{diaz2020non}. Having introduced $\psi_t$, the sequentially doubly robust estimator for the case of two time points can be computed based on the following algorithm:
\begin{enumerate}
    \item Fit a regression model $\hat m_2(a_2, h_2)=\hat E(Y\mid A_2=a_2, H_2=h_2)$. That is, regress the outcome $Y$ on  variables $A_2$ and $H_2$. This is the same as Step 1 above.
        \item Predict the outcome under the hypothetical intervention. That is, compute the prediction $\hat m_2(\dd_2(A_2, A_1), H_2)$ for every unit in the sample. This is the same as Step 2 above.
    \item Fit a model $\hat g_2(a_2\mid h_2)=\hat P(A_2=a_2\mid H_2=h_2)$
    \item Compute the estimated pseudo-outcome $\hat \psi_2(Z)$ using $\hat m_2$ and $\hat g_2$.
    \item Fit a regression model $\hat m_1(a_1, h_1)=\hat E(\hat\psi_2(Z)\mid A_1=a_1, H_1=h_1)$. That is, regress the pseudo-outcome $\hat \psi_2(Z)$ on  variables $A_1$ and $H_1$.
        \item Predict the pseudo-outcome under the hypothetical intervention. That is, compute the prediction $\hat m_1(\dd_1(A_1, A_0), H_1)$ for every unit in the sample.
        \item Fit a model $\hat g_1(a_1\mid h_1)=\hat P(A_1=a_1\mid H_1=h_1)$
    \item Compute the estimated pseudo-outcome $\hat \psi_1(Z)$ using $\hat m_2$, $\hat g_2$, $\hat m_1$, and $\hat g_1$.
    \item  The final estimate of the mean outcome under the hypothetical intervention is the average over the entire sample of $\hat \psi_1(Z)$.
\end{enumerate}

In addition to using a doubly robust transformation $\psi_t$, the estimator used in this paper uses cross-fitting, a technique that generally improves the finite sample performance of the estimator. In cross-fitting, the data is partitioned into equally sized prediction sets, with the corresponding training sets being defined as the rest of the sample. The regressions are fit independently on each training set, and then used to predict in the corresponding prediction set. This algorithm is implemented in the lmtp R package.\cite{williams2020lmtp} 

This procedure can be extended to multiple time points, and arbitrary regression algorithms may be used. We provide commented code that implements this estimator in our data: \url{https://github.com/kararudolph/code-for-papers/blob/master/NALeffects.R}. In addition, other tutorials are available: \url{https://github.com/nt-williams/lmtp}.

\end{document}